\documentclass[cameraready]{Interspeech}
\usepackage[T1]{fontenc}
\usepackage{amsmath} 
\usepackage{amsfonts}   
\usepackage{amssymb}    
\usepackage{graphicx}   
\usepackage{booktabs}   
\usepackage{array}      
\usepackage{tabularx}   
\usepackage{hyperref}

\title{BridgeCodec: Mamba Enhanced Neural Audio Codec with Schrödinger Bridge at Low Bitrate}

\author[affiliation={1}]{Zijian}{Lin}
\author[affiliation={2}]{Jing}{Yang}
\author[affiliation={2}]{Jinghao}{Luo}
\author[affiliation={2}]{Zhuo}{Wang}
\author[affiliation={2}]{Fan}{Fan}
\author[affiliation={1}, correspondingauthor]{Zhiyong}{Wu}

\address{
$^1$ Shenzhen International Graduate School, Tsinghua University, China \\
$^2$ Central Media Technology Institute, Huawei
}

\email{linzj24@mails.tsinghua.edu.cn, zywu@sz.tsinghua.edu.cn}


\usepackage{comment}

\begin{document}

\maketitle

\begin{abstract}
Neural audio codecs achieve remarkable performance but operate as closed ecosystems with rigidly co-trained encoder-decoder pairs, severely hindering interoperability. To resolve this, we propose BridgeCodec, a versatile framework that decouples these pairs by translating latent representations between mismatched, frozen endpoints. We formulate this cross-codec translation as a Schrödinger Bridge optimal transport problem, enabling robust mapping between disparate distributions without predefined noise priors. To efficiently capture long-range speech temporal dependencies, we integrate a Mamba-enhanced U-Net architecture. We validate BridgeCodec on an extreme deployment scenario: mapping a lightweight 8~kHz source encoder directly to a 48~kHz target decoder at an ultra-low bitrate of 1~kbps. Experimental results demonstrate that BridgeCodec achieves superior wideband reconstruction and high perceptual quality, successfully bridging heavily bottlenecked, heterogeneous codecs.\footnote{Demo: https://thuhcsi.github.io/interspeech2026-BridgeCodec/}
\end{abstract}

\section{Introduction}
\label{sec:intro}
Neural audio codecs (NACs) have revolutionized audio processing by achieving a unified framework for efficient compression and high-fidelity reconstruction \cite{zeghidour2021soundstream}. Beyond traditional transmission, these systems now serve as critical intermediaries in downstream speech synthesis systems. State-of-the-art NACs typically employ Residual Vector Quantization (RVQ) \cite{zeghidour2021soundstream, defossez2023high, kumar2023high,10768970} to capture fine-grained acoustic details. Recent advancements further branch into two paradigms: integrating semantic information via multi-codebook designs to enhance LLM prediction \cite{ZhangZLZQ24, defossez2024moshi,ye2025codec,gong2025xy}, and exploring single-layer VQ to push the boundaries of ultra-low bitrates \cite{OordVK17,Ji00C0Z0C0LZY0J25, xin2024bigcodec,10446556}.

Despite their remarkable performance, current NACs operate as closed ecosystems relying on strictly co-trained encoder-decoder pairs. This coupling creates a severe "Interoperability Gap" across heterogeneous networks, preventing communication between devices with codecs deployed independently for with asymmetric hardware capabilities or varying compute constraints. Since simultaneously updating these structurally frozen endpoints may be practically infeasible and risks systemic failures, direct communication is fundamentally broken. To resolve this, we propose \textit{BridgeCodec}, a versatile framework that completely decouples codec pairs by directly translating highly compressed latent representations between any mismatched, frozen endpoints. To demonstrate its robustness in extreme deployment scenarios, we rigorously validate BridgeCodec on a highly asymmetric testbed: mapping a lightweight 8~kHz encoder to a 48~kHz decoder at an ultra-low 1~kbps. This confirms its capability to seamlessly bridge structurally heterogeneous latents and fully activate the generative priors of the target decoder.

To decouple encoder and decoder, we draw inspiration from the recent success of the Schrödinger Bridge (SB) \cite{schrodinger1932theorie, de2021diffusion} in diverse speech tasks, including generation \cite{chen2023schrodinger}, enhancement \cite{han25b_interspeech, yang2025schr}, bandwidth extension \cite{li2025bridge, li2025audio} and speech vocoder \cite{lei2025bridgevoc}, where it has been extensively validated to outperform flow matching \cite{lipman2023flow} and standard diffusions \cite{song2019generative, 0011SKKEP21}. Therefore, we formulate our latent transformation as an SB optimal transport problem. Crucially, unlike standard score-based and flow models that require predefined noise priors, SB natively maps highly compressed, disparate latent spaces by adaptively learning optimal transport between empirical distributions. Moreover, capturing the long-range temporal dependencies intrinsic to speech (e.g., prosody, intonation, and phonetic coherence) is critical for high-fidelity reconstruction. Extensive recent literature \cite{miyazaki24_interspeech, 10985910, 10889111} has firmly established the superiority of Mamba \cite{gu2023mamba}  in modeling extended speech contexts. Based on these findings, we integrate Mamba into a U-Net backbone \cite{ronneberger2015u}. This tailored architecture enables BridgeCodec to capture global temporal dynamics with linear complexity, strategically bypassing the localized receptive fields of convolutions and the quadratic computational bottlenecks of Transformers. Finally, we maximize perceptual quality through a two-stage training strategy: structural latent-space alignment followed by audio-domain refinement. Experimental results demonstrate that BridgeCodec successfully bridges the mismatched codecs under stringent low-bitrate constraints. Remarkably, it sustains high perceptual quality even with a single function evaluation (1 NFE), empirically validating the structural directness of our chosen SB transport path and robustness of the framework.

Our contributions are summarized as follows. First, we propose \textit{BridgeCodec}, a SB based framework that decouples encoder-decoder to enable direct speech communication between mismatched endpoints. Second, we introduce a Mamba-enhanced U-Net architecture tailored to efficiently capture long-range speech dependencies within compressed representations. Third, leveraging a two-stage training strategy, we demonstrate that BridgeCodec achieves high-fidelity 48~kHz wideband reconstruction from 8~kHz bottlenecked priors at 1~kbps.

\begin{figure*}[h]
    \centering
    \includegraphics[width=0.75\linewidth]{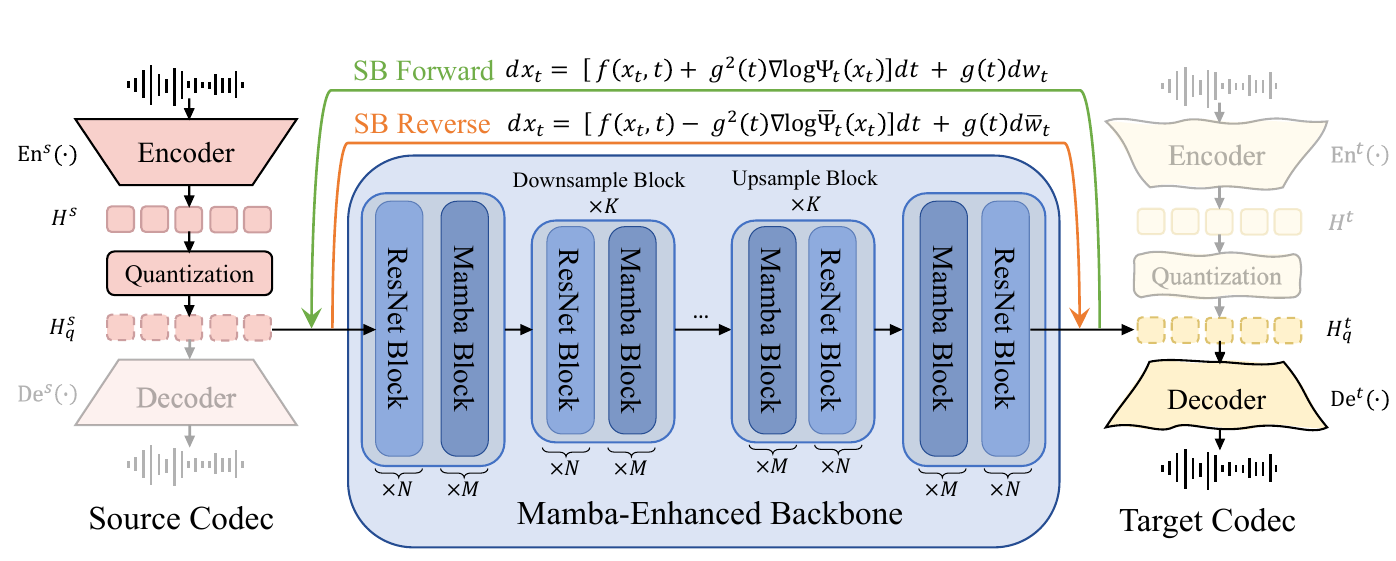}
    \caption{The BridgeCodec framework consists of \(\text{En}^s(\cdot)\), Mamba-enhanced backbone, and \(\text{De}^t(\cdot)\). The source codec is marked with straight lines and the target codec with wavy lines, indicating they can use different model architectures and quantization methods.}
    \label{fig:overall}
    \vspace{-15pt}
\end{figure*}

\section{Methodology}
\label{sec:metho}

\subsection{Overall Architecture}

Neural audio codecs (NACs) typically compress an audio waveform $X \in \mathbb{R}^{T_X}$ via an encoder $\text{En}(\cdot)$ into a latent representation $H \in \mathbb{R}^{D_c \times T_H}$, where $D_c$ and $T_H$ denote the channel dimension and frame length, respectively. A quantizer with $L$ vector quantization (VQ) layers then maps $H$ into discrete tokens $Q \in \mathbb{R}^{L \times T_H}$. The embeddings corresponding to these tokens are summed across all $L$ layers to reconstruct the quantized continuous feature $H_q \in \mathbb{R}^{D_c \times T_H}$. Finally, a decoder $\text{De}(\cdot)$ reconstructs the audio $X_q$ from $H_q$.

To achieve interoperability between heterogeneous codecs, BridgeCodec fundamentally decouples the source encoder $\text{En}^s(\cdot)$ and the target decoder $\text{De}^t(\cdot)$. We formulate an intermediary latent translator using the Schrödinger Bridge (SB) \cite{schrodinger1932theorie, de2021diffusion}. As illustrated in Fig.~\ref{fig:overall}, a Mamba-enhanced backbone translates the source latent feature $H_q^s$ directly into the target latent feature $H_q^t$. Because this transformation operates entirely in the latent space, the source and target codecs can maintain entirely different architectures, dimensions, and quantization paradigms. During deployment, frozen $\text{En}^s(\cdot)$, SB translator, and frozen $\text{De}^t(\cdot)$ seamlessly compose a unified and flexible asymmetric communication pipeline.

In this work, we validate BridgeCodec on a rigorous asymmetric testbed: connecting a lightweight, low-bitrate narrowband (8~kHz) source encoder to a high-fidelity wideband (48~kHz) target decoder. In this demanding setup, the SB-driven translator is tasked with not only aligning structurally mismatched latent dimensions but also performing latent-space bandwidth extension. By effectively modeling and compensating for the high-frequency semantics missing in the heavily bottlenecked $H_q^s$ prior, BridgeCodec robustly recovers 48~kHz wideband audio without requiring any structural modifications to the endpoints.

\subsection{Schr\"odinger Bridges}

The core challenge in BridgeCodec is establishing a direct mapping between the source latent $H_q^s$ and the target latent $H_q^t$. Standard diffusions are mathematically ill-suited for this task, as they constrain the terminal distribution to a Gaussian noise prior ($p_T = \mathcal{N}(0, \mathbf{I})$), which destructively overwrites the structural speech priors preserved in $H_q^s$. Inspired by recent success of SB in speech tasks \cite{chen2023schrodinger,han25b_interspeech,yang2025schr,li2025bridge,li2025audio,lei2025bridgevoc}, we bypass this limitation by formulating the latent translation as a SB optimal transport problem. Because SB adaptively learns transitions between arbitrary empirical distributions without predefined noise priors, it uniquely qualifies to directly and robustly bridge two structurally mismatched, compressed latent spaces.

Formally, SB finds an optimal path measure \(p \in \mathcal{P}_{[0,T]}\) that minimizes the Kullback-Leibler divergence with a reference path \(p_{\text{ref}}\), while satisfying boundary constraints \(p_0 = p_{\text{target}}\) and \(p_T = p_{\text{source}}\):
\begin{equation}
\label{equ:kl_bridge}
    \min_{p \in \mathcal{P}_{[0,T]}} D_{\text{KL}}(p \parallel p_{\text{ref}}) \quad \text{s.t.} \quad p_0 = p_{\text{target}},\ p_T = p_{\text{source}},
\end{equation}
where \(p_{\text{source}}\) and \(p_{\text{target}}\) are the two distributions to be connected. The path measure solution to Equation~\ref{equ:kl_bridge} is a pair of forward-backward stochastic differential equations (SDEs):
\begin{align}
&dx_t = \left[ f(x_t, t) + g^2(t) \nabla \log \Psi_t(x_t) \right] dt + g(t) dw_t,\\
&dx_t = \left[ f(x_t, t) - g^2(t) \nabla \log \bar{\Psi}_t(x_t) \right] dt + g(t) d\bar{w}_t,
\end{align}
where \( f(x_t, t) \) and \( g(t) \) are inherited from the reference SDE defining \( p_{\text{ref}} \); \( dw_t \) and \( d\bar{w}_t \) denote distinct Wiener processes capturing stochasticity. 
The forward and backward evolution are characterized by distinct nonlinear drift terms \(+g^2(t) \nabla \log \Psi_t(x_t)\) and \(-g^2(t) \nabla \log \bar{\Psi}_t(x_t)\). 

The marginal distribution of \(x_t\) is given by \(p_t = \Psi_t \bar{\Psi}_t\) for any \(t \in [0, T]\).

SB has tractable closed-form solutions under linear drift \(f(x_t)=f(t)x_t\) and Gaussian boundaries, avoiding general computational intractability~\cite{chen2023schrodinger}. Let \(p_0=\mathcal{N}(x_0,\epsilon_0^2\mathbf{I})\) and \(p_T=\mathcal{N}(x_T,\epsilon_T^2\mathbf{I})\), with \(\epsilon_T=e^{\int_0^T f(\tau)d\tau}\epsilon_0\). As \(\epsilon_0\to0\), the forward/backward terms \(\Psi_t,\bar{\Psi}_t\) can be formulated as 
$\bar{\Psi}_t=\mathcal{N}(\alpha_tx_0,\alpha_t^2\sigma_t^2\mathbf{I}),\quad\Psi_t=\mathcal{N}(\bar{\alpha}_tx_T,\alpha_t^2\bar{\sigma}_t^2\mathbf{I})$,
where \(\alpha_t=e^{\int_0^t f(\tau)d\tau}\), \(\sigma_t^2=\int_0^t g^2(\tau)/\alpha_\tau^2d\tau\), \(\bar{\alpha}_t=\alpha_t\alpha_T^{-1}\), \(\bar{\sigma}_t^2=\sigma_T^2-\sigma_t^2\). The marginal \(p_t=\Psi_t\bar{\Psi}_t\) is thus formulated as:
\begin{equation}
    p_t=\mathcal{N}(w_0(t)x_0+w_T(t)x_T,\alpha_t^2\sigma_t^2\bar{\sigma}_t^2/\sigma_T^2\mathbf{I}),
\end{equation}
where, \(w_0(t)=\alpha_t\bar{\sigma}_t^2/\sigma_T^2\), \(w_T(t)=\bar{\alpha}_t\sigma_t^2/\sigma_T^2\) are time-varying weights balancing initial/terminal mean contributions. 

In the context of our work, the terminal distribution mean \( x_T \) and the initial distribution mean \( x_0 \) correspond to the source codec feature $H_q^s$ and the target codec feature $H_q^t$, respectively. Optimization uses a bridge loss predicting \( x_0 \) from sampled \( x_t \sim p_t \):
\begin{equation}
\label{equ:bridge_loss}
    \mathcal{L}_{\text{bridge}} = \mathbb{E}_{x_T \sim p_T, x_0 \sim p_0} \mathbb{E}_t \left[ \lVert x_\theta(x_t, t, x_T) - x_0 \rVert_2^2 \right],
\end{equation}
where $ x_\theta(\cdot) $ denotes the parameterized prediction model.


\subsection{Mamba-Enhanced Backbone}

While standard U-Nets excel at extracting localized structural patterns from acoustic latents, they inherently lack the receptive field to capture long-range temporal correlations critical to speech, such as prosodic fluctuations and phonetic coherence. To overcome this localized bias, we incorporate Mamba \cite{gu2023mamba}, a selective State Space Model, into our latent translation backbone. Using input-dependent discrete recurrences, Mamba effectively models global sequential dynamics in linear time. Crucially, recent speech processing literature \cite{miyazaki24_interspeech, 10985910, 10889111} has demonstrated that Mamba matches or even exceeds the generative fidelity of Transformers while maintaining significantly faster inference speeds. This avoids the quadratic computational bottlenecks of attention mechanisms while perfectly complementing the spatial feature processing capabilities of U-Nets.

Architecturally, we interleave Mamba modules within the $K$ resolution levels of the U-Net, as shown in Fig.~\ref{fig:overall}. At each encoder downsampling stage, the input latent $H_q^s \in \mathbb{R}^{B \times 1 \times D_c \times T_H}$ is first processed by $N$ convolutional residual blocks. This local feature extraction expands the representation to an intermediate tensor $H_{\text{conv}} \in \mathbb{R}^{B \times C \times D_r \times T_H}$, where $C$ and $D_r$ denote the channel and hidden dimensions, respectively. To establish a global temporal context, $H_{\text{conv}}$ is flattened along its spatial dimensions into a sequence $\mathbb{R}^{B \times T_H \times D_m}$ (with $D_m = C \times D_r$) and routed through $M$ consecutive Mamba layers. The globally-enriched output is subsequently reshaped back to its original topology $\mathbb{R}^{B \times C \times D_r \times T_H}$ to ensure seamless propagation to the next resolution level. The decoder strictly mirrors this Mamba-enhanced hierarchy, ensuring synchronized local-global feature alignment throughout the Schrödinger Bridge transport path.

\begin{table*}[htbp]
  \centering
  \caption{Main results of BridgeCodec on the testset. All audio is resampled to 48~kHz for STOI, STFT and MEL. And due to evaluation model constraints, all audio is resampled to 16~kHz for DNSMOS, WER and SIM. MOS represents the subjective Mean Opinion Score for naturalness and speech quality with 95\% confidence intervals.}
  \footnotesize
  \renewcommand\arraystretch{0.8}
  \begin{tabular}{l|ccc|ccc|c} 
    \toprule 
    \textbf{Model} & \textbf{STOI} $\uparrow$ & \textbf{STFT} $\downarrow$ & \textbf{MEL} $\downarrow$ & \textbf{DNSMOS} $\uparrow$  & \textbf{WER} $\downarrow$ & \textbf{SIM} $\uparrow$ & \textbf{MOS} $\uparrow$ \\
    \midrule
    target codec   & 0.96 & 1.63 & 1.05 & 3.26 & 1.70  & 0.99 & 4.24 $\pm$ 0.10 \\
    source codec    & 0.23 & 12.24 & 10.01 & 3.11 & 1.92 & 0.81 & 3.80 $\pm$ 0.11 \\
    \midrule 
    BridgeCodec$_\text{UNet}$ & 0.86 & 2.27 & 1.74 & 3.09  & 2.16  & 0.93 & - \\
    BridgeCodec       & \textbf{0.88}   & 2.22      & \textbf{1.65}   & \textbf{3.17}  & 2.11  & \textbf{0.96} & \textbf{4.12 $\pm$ 0.11} \\
    \quad w/o Two stage training & 0.87   & \textbf{2.19}      & 1.68   & 3.15  & \textbf{2.04}  & 0.95 & - \\
    \bottomrule 
  \end{tabular}
  \label{tab:audio_codec_comparison}
  \vspace{-10pt}
\end{table*}

\subsection{Two-Stage Training Strategy}
While optimizing the SB objective ($\mathcal{L}_{\text{bridge}}$) ensures strict optimal transport in the latent space, relying solely on latent alignment can lead to sub-optimal perceptual quality, as the model remains blind to the acoustic characteristics of the actual decoded waveforms. To align latent structural accuracy with audio-domain fidelity, we employ a two-stage training strategy. First, the SB model is trained exclusively in the highly compressed latent domain using Equation \ref{equ:bridge_loss} to establish a robust transport path between the heterogeneous codecs. Then, at the second stage, to inject real-world acoustic constraints, the pre-trained SB translator is cascaded with the frozen target decoder $\text{De}^t(\cdot)$ to reconstruct the time-domain waveform $\hat{X}$. The translator is then fine-tuned against the ground-truth wideband audio $X$ using a joint objective:
\begin{equation}
\mathcal{L}_{\text{total}} = \lambda_{\text{bridge}} \mathcal{L}_{\text{bridge}} + \lambda_t \mathcal{L}_\text{time} + \lambda_{\text{mel}} \mathcal{L}_{\text{mel}},
\end{equation}
where $\mathcal{L}_\text{time} = \lVert X - \hat{X} \rVert_1$ is the time-domain L1 loss, and $\mathcal{L}_{\text{mel}} = \lVert S(X) - S(\hat{X}) \rVert_1$ is the frequency-domain mel-spectrogram L1 loss \cite{kumar2023high}, with $S(\cdot)$ denoting the mel-extraction function. The scalars $\lambda_{\text{bridge}}$, $\lambda_t$, and $\lambda_{\text{mel}}$ balance the respective loss terms.

\section{Experiments}
\label{sec:exset}
\subsection{Dataset}
We conducted experiments on the clean speech dataset from the ICASSP 2022 Deep Noise Suppression (DNS) Challenge \cite{DubeyGCAMBETYGA22}. 
This dataset supports fullband 48~kHz audio processing, with a total duration of approximately 2,425 hours and around 1.09 million clips covering six languages. 
10,000 clips (around 1\% of the dataset) are designated as the validation set and another 10,000 clips as the test set. The remaining majority of clips are used as the training set.

\subsection{Implementation Details and Baselines}
We trained a source codec and a target codec for 8~kHz and 48~kHz speech separately using open-source AudioCraft.\footnote{https://github.com/facebookresearch/audiocraft} The source codec uses ratios [2, 4, 5, 8] to produce 128-dimensional latents at 25 Hz, with 4-layer RVQ and 1024 codebooks per layer, achieving 1~kbps with 2.8 MB parameters. The target codec uses ratios [2, 4, 5, 6, 8] to produce 128-dimensional latents at 25 Hz, with 24-layer RVQ and 1024 codebooks per layer, achieving 6~kbps with 134 MB parameters. 
The SB model in $\text{BridgeCodec}$ uses a U-Net backbone with 4 resolution levels and 2× downsampling per level, each followed by 4 residual blocks and 5 Mamba layers whose hidden size is 128 for long-range temporal modeling. 
It was trained in two stages, first with learning rate 5e-5, batch size 32 for 100k steps, and then with learning rate 5e-6, batch size 8 for 10k steps. We used VP noise schedule following the setting of Bridge-TTS \cite{chen2023schrodinger} and SDE sampler with 10 NFE for inference. The number of function evaluations (NFE) refers to the count of forward passes of the neural network during sampling, directly related to computational cost. For comparative evaluation, we trained an SB model based on NCSN++~\cite{0011SKKEP21} with the same parameters and configuration as the baseline but excluding the Mamba layers, named $\text{BridgeCodec}_\text{UNet}$.

\subsection{Evaluation Metrics}

We adopt several metrics to evaluate the performance of Bridgecodec. STOI \cite{taal2011algorithm} is an intrusive metric that assesses speech intelligibility. Word Error Rate (WER) quantifies audio intelligibility by calculating the discrepancies between transcriptions and original texts. We only calculate WER on the English subset of the test set and transcribe the speeches through a Whisper-large V3 \cite{radford2023robust}. Speaker similarity (SIM) reflects the ability of the codec to preserve speaker characteristics. We use WavLM-base-sv \cite{chen2022wavlm} to extract the speaker embeddings of the reconstructed and ground-truth waveforms, and calculate the cosine similarity between them to measure SIM. DNSMOS \cite{reddy2022dnsmos} is a widely adopted metric for evaluating objective speech quality. Mel distance (MEL) \cite{kumar2023high} calculates the distance between log mel spectrograms of the reconstructed and ground-truth waveforms. The multiscale STFT distance (STFT) \cite{kumar2023high} computes the distance between log magnitude spectrograms of the reconstructed and ground-truth waveforms. To evaluate perceptual fidelity from a human perspective, we conduct a subjective Mean Opinion Score (MOS) test on the naturalness of the reconstruction speech. A panel of 30 fluent listeners evaluates the reconstructed speech quality of 15 randomly sampled test-set clips on a standard 5-point scale with 1-point intervals.

\section{Results}
\label{sec:res_ana}

Table \ref{tab:audio_codec_comparison} presents the comprehensive evaluation of BridgeCodec against the mismatched baselines. The target codec (48~kHz, 6~kbps) serves as the upper bound for high-fidelity reconstruction, while the source codec (8~kHz, 1~kbps) acts as the lower bound, exhibiting severely degraded performance across all metrics due to its narrowband constraint and ultra-low bitrate.

\subsection{Objective Results}

BridgeCodec successfully closes the massive interoperability gap between these endpoints. Without altering the 1~kbps transmission bitrate, it dramatically improves upon the source codec's baseline. Notably, STOI leaps from 0.23 to 0.88, and MEL drops from 10.01 to 1.65, indicating the improvement of speech intelligibility. Most impressively, BridgeCodec effectively reconstructs speaker characteristics during the extreme bandwidth extension, elevating SIM from 0.81 to 0.96, closely approaching the target codec's upper bound of 0.99. While the WER sees a marginal increase compared to the bounds (2.11 vs. 1.92), this is an expected outcome given the massive generative hallucination required to map 8~kHz latents to 48~kHz wideband audio. The source encoder captures only narrowband, and the bridge performs generative extension. Generated high-frequency content may slightly perturb Whisper, leading to increased WER. Despite this, the semantic content and  speech quality remains highly well-preserved.

\subsection{Subjective Results}

The objective improvements are further strongly corroborated by our subjective MOS evaluation. BridgeCodec achieves a remarkable MOS of 4.12 $\pm$ 0.11, significantly outperforming the source codec (3.80 $\pm$ 0.11) and demonstrating that the generated 48~kHz audio sounds highly natural to human listeners. This confirms that BridgeCodec does not merely interpolate signals, but successfully activates the generative priors of the target decoder to produce perceptually pleasing wideband speech.

\subsection{Ablation Studies}

To validate our architectural and methodological designs, we conduct ablation studies on the Mamba backbone and the training strategy, as shown in Table~\ref{tab:audio_codec_comparison}.

\noindent \textbf{Effectiveness of Mamba.} The proposed model consistently outperforms the baseline BridgeCodec$_{\text{UNet}}$ across all metrics. This confirms that while standard convolutions capture localized spatial correlations in the latent space, Mamba's selective state space mechanism is essential for modeling long-range global temporal dependencies (such as prosody and pitch variations), thereby reducing semantic distortions.

\noindent \textbf{Impact of Two-Stage Training.} Comparing BridgeCodec with its "w/o Two-stage training" variant reveals that involving the two-stage training strategy enhances performance across most metrics, with only minor trade-offs observed in STFT and WER, suggesting that latent domain training alone may more effectively preserve semantic information. However, second-stage training is critical for improving perceptual quality and speaker similarity, allowing BridgeCodec to achieve higher DNSMOS and SIM.

\noindent \textbf{Sensitivity to Inference Steps.} To investigate the impact of inference step on reconstruction performance, we conduct an experiment on the number of function evaluations (NFE) in the Schrödinger Bridge sampler, as shown in Table~\ref{tab:ablation}. Typically, fewer NFEs translate to lower inference latency and higher efficiency, but this compromises reconstruction performance due to insufficient iterative refinement of latent features \cite{li2025bridge}. However, as illustrated in Table \ref{tab:ablation}, BridgeCodec demonstrates a remarkable robustness to reduced NFE. Specifically, when NFE is minimized to 1, the model achieves the best performance in STOI, STFT, MEL, and DNSMOS.

This one-step capability suggests that the Schrödinger Bridge in our framework learns an exceptionally straight and efficient transport path between the source and target latents. While increasing NFE to 10 provides a marginal gain in speaker similarity (SIM) through iterative refinement, the 1-NFE configuration provides a superior balance of acoustic fidelity and structural alignment. This makes BridgeCodec a highly promising candidate for low-latency communication where minimal inference iterations are critical.

\begin{table}[htbp]
  \centering
  \caption{The results with different number of function evaluations.}
  \footnotesize
  \renewcommand\arraystretch{0.8}
  \begin{tabular}{ccccccc} 
    \toprule 
    \textbf{NFE} & \textbf{STOI} & \textbf{STFT} & \textbf{MEL} & \textbf{DNSMOS} & \textbf{WER}  & \textbf{SIM} \\
    \midrule
    10  & 0.88 & 2.22 & 1.65 & 3.17  & \textbf{2.11}  & \textbf{0.96}  \\
    5  & 0.88 & 2.16 & 1.63 & 3.17 & 2.37 & 0.95 \\
    1  & \textbf{0.89} & \textbf{2.13} & \textbf{1.58} & \textbf{3.18} & 2.39 & 0.95 \\
    \bottomrule 
  \end{tabular}
  \vspace{-20pt}
  \label{tab:ablation}
\end{table}

\subsection{Deployment Efficiency and Universal Interoperability}

Beyond individual reconstruction metrics, BridgeCodec establishes a new paradigm for codec interoperability. While our experiments utilize the challenging 8~kHz to 48~kHz transformation as a primary testbed, the core value of BridgeCodec lies in its ability to perform heterogeneous latent translation between structurally mismatched manifolds. By decoupling the source encoder from the specific requirements of target decoder, BridgeCodec effectively serves as a universal translation framework. This allows diverse audio codecs—regardless of their sampling rates, bitrates, or internal architectures—to communicate within a unified framework. Such a capability is crucial for breaking the closed-loop constraints of current NACs, enabling seamless high-fidelity audio flow across fragmented, resource-constrained network ecosystems.

\section{Conclusion}
\label{sec:concl}

We presented BridgeCodec, a framework for universal audio codec interoperability that decouples rigidly paired encoders and decoders. By integrating Schrödinger Bridge for latent transport and a Mamba-enhanced backbone for temporal modeling, BridgeCodec achieves high-fidelity 48~kHz reconstruction from 8~kHz inputs at an ultra-low 1~kbps. Our two-stage training strategy effectively balances structural alignment with perceptual naturalness, supporting robust single-step inference. This work provides a scalable solution for high-quality and seamless audio communication across heterogeneous and resource-constrained network ecosystems.

\section{Generative AI Use Disclosure}
During the preparation of this manuscript, the authors used generative AI tools exclusively for the purpose of language editing and manuscript polishing to improve readability. These tools were not used to generate any core scientific ideas, experimental data, or technical contributions. All authors have thoroughly reviewed and approved the final version of the manuscript, and assume full responsibility for the integrity and entirety of its content.

\section{Acknowledgement}
This work was supported by National Natural Science Foundation of China (62076144).

\bibliographystyle{IEEEtran}
\bibliography{mybib}

\end{document}